\documentclass[12pt]{article} 
\pdfoutput=1

\usepackage[utf8]{inputenc}
\usepackage[T1]{fontenc} 
\usepackage{lmodern}
\usepackage[english]{babel}           
\usepackage{cancel}
\usepackage[pdftex]{graphicx}
\usepackage{epsfig}
\usepackage{graphicx}
\usepackage{comment}
\usepackage{latexsym}
\usepackage{hyperref}
\usepackage{amsmath}
\usepackage{mathrsfs}
\usepackage[usenames, dvipsnames]{color}
\usepackage{amsbsy}
\usepackage{amssymb}
\usepackage{amsthm}
\usepackage{amsfonts}
\usepackage{cite}
\usepackage{upgreek}

\usepackage{xcolor}
\usepackage{color}
\usepackage{mathtools}
\usepackage{caption} 
\usepackage{subcaption} 
\usepackage{euscript} 
\usepackage{float}
\usepackage{diagbox}
\usepackage[title]{appendix}
\usepackage{tabularx}

\newcommand{\M}{{\cal M}}
\newcommand{\N}{{\cal N}}
\newcommand{\T}{{\cal T}}
\newcommand{\V}{{\cal V}}

\newcommand{\F}{{\cal F}}
\newcommand{\D}{{\cal D}}
\newcommand{\J}{{\cal J}}
\renewcommand{\i}{{\rm i}}
\newcommand{\f}{{\rm f}}
\newcommand\q{{\rm q}}
\newcommand\Q{{\rm Q}}

\newcommand\Vol{{\rm Vol}}
\newcommand\Diff{{\rm Diff}}

\renewcommand\d{{\rm d}}

\DeclareMathOperator\sign{sign}

\newcommand{\be}{\begin{equation}}
	\newcommand{\ee}{\end{equation}}
\newcommand{\dis}{\displaystyle}
\renewcommand{\thefootnote}{\fnsymbol{footnote}}
\newcommand{\eq}[1]{\eqref{#1}}
\newcommand{\Eq}[1]{Eq.~\eqref{#1}}
\newcommand{\Eqs}[1]{Eqs.~\eqref{#1}}
\newcommand{\Refe}[1]{Ref.~\cite{#1}}
\newcommand{\Refs}[1]{Refs.~\cite{#1}}
\newcommand{\Sect}[1]{Sect.~\ref{#1}}

\newcommand{\Appendix}[1]{Appendix~\ref{#1}}

\newcommand{\R}{\mathbb{R}}
\newcommand{\Z}{\mathbb{Z}}

\renewcommand{\O}{{\cal O}}

\newcommand{\ie}{{\it i.e.} }
\newcommand{\eg}{{\it e.g.} }

\newcommand{\apriori}{{\it a priori} }
\newcommand{\where}{\mbox{where}}
\newcommand{\with}{\mbox{with}}
\newcommand{\when}{\mbox{when}}
\renewcommand{\and}{\mbox{and}}

\newcommand{\esps}{\phantom{\!\!\!\overset{|}{a}}}
\newcommand{\esp}{\phantom{\!\!\overset{\displaystyle |}{|}}}
\newcommand{\espD}{\phantom{\!\!\underset{\displaystyle |}{\cdot}}}

\newcommand{\bm}{\boldmath} 

\usepackage[
a4paper,
top=2.5cm,
bottom=2.5cm,
left=2.5cm,
right=2.5cm,
headheight=14pt, 
]{geometry}

\catcode`\@=11
\def\marginnote#1{}
\newcount\hour
\newcount\minute
\newtoks\amorpm
\hour=\time\divide\hour by60 \minute=\time{\multiply\hour by60
	\global\advance\minute by-\hour}
\edef\standardtime{{\ifnum\hour<12 \global\amorpm={am}%
		\else\global\amorpm={pm}\advance\hour by-12 \fi
		\ifnum\hour=0 \hour=12 \fi
		\number\hour:\ifnum\minute<10 0\fi\number\minute\the\amorpm}}
\edef\militarytime{\number\hour:\ifnum\minute<10 0\fi\number\minute}
\def\draftlabel#1{{\@bsphack\if@filesw {\let\thepage\relax
			\xdef\@gtempa{\write\@auxout{\string
					\newlabel{#1}{{\@currentlabel}{\thepage}}}}}\@gtempa
		\if@nobreak \ifvmode\nobreak\fi\fi\fi\@esphack}
	\gdef\@eqnlabel{#1}}
\def\@eqnlabel{}
\def\@vacuum{}
\def\draftmarginnote#1{\marginpar{\raggedright\scriptstyle\tt#1}}
\def\draft{\oddsidemargin -.2truein
	\def\@oddfoot{\sl preliminary draft \hfil
		\rm\thepage\hfil\sl\today\quad\militarytime}
	\let\@evenfoot\@oddfoot \overfullrule 3pt
	\let\label=\draftlabel
	\let\marginnote=\draftmarginnote
	\def\@eqnnum{(\theequation)\rlap{\kern\marginparsep\tt\@eqnlabel}%
		\global\let\@eqnlabel\@vacuum}  }
\def\thebibliography#1{
	\vskip 0.5cm \centerline{\bf \Large References}
	\list{
		[\arabic{enumi}]}{\settowidth\labelwidth{[#1]}
		\leftmargin\labelwidth
		\advance\leftmargin\labelsep
		\usecounter{enumi}}
	\def\newblock{\hskip .11em plus .33em minus .07em}
	\sloppy\clubpenalty4000\widowpenalty4000
	\sfcode`\.=1000\relax}

\renewcommand{\theequation}{\arabic{section}.\arabic{equation}}
\renewcommand{\section}{\setcounter{equation}{0}\@startsection
	{section}{1}{0mm}{-\baselineskip}{0.5\baselineskip} {\normalfont\Large\bfseries}}
\renewcommand{\subsection}{\@startsection
	{subsection}{2}{0mm}{-\baselineskip}{0.5\baselineskip} {\normalfont\large\bfseries}}
\renewcommand{\subsubsection}{\@startsection
	{subsubsection}{3}{0mm}{-\baselineskip}{0.5\baselineskip}
	{\normalfont\normalsize\slshape}}

\begin{document}
	
	
	\begin{titlepage}
		\begin{flushright}
			CPHT-RR053.102025, December 2025
			\vspace{0.0cm}
		\end{flushright}
		\vspace{6mm}
		
		\begin{centering}
			{\bm\bf  \Large Ordering-Independent Wheeler--DeWitt Equation \vspace{.2cm}
			
			for Flat Minisuperspace Models}

            


			
			\vspace{9mm}
			
			{\bf Victor Franken,$^{1,2}$\footnote{victor.franken@ugent.be} Eftychios Kaimakkamis,$^3$\footnote{kaimakkamis.eftychios@ucy.ac.cy} Herv\'e Partouche$^1$\footnote{herve.partouche@polytechnique.edu}\\ \vspace{0.1cm} and Nicolaos Toumbas$^3$\footnote{nick@ucy.ac.cy}}
			
			\vspace{4mm}

			$^1$ {\em CPHT, CNRS, Ecole polytechnique, IP Paris, \\F-91128 Palaiseau, France} 
			
			$^2$ {\em Department of Physics and Astronomy, Ghent University,\\
				Krijgslaan, 281-S9, 9000 Gent, Belgium}
			
			$^3$ {\em Department of Physics, University of Cyprus, \\Nicosia 1678, Cyprus}

		\end{centering}
		\vspace{0.2cm}
		$~$\\
		\centerline{\bf\Large Abstract}\\
		\vspace{-0.8cm}
		
		\begin{quote}
		We consider minisuperspace models with two-derivative kinetic terms, assuming a flat target space and a closed Universe. We show that, upon canonical quantization of the Hamiltonian, only a restricted class of operator orderings is compatible with the path-integral formulation. Remarkably, these orderings are physically equivalent to all orders in~$\hbar$.
More precisely, each choice of path-integral measure in the definition of the wavefunction path integral uniquely determines an operator ordering, and hence a corresponding Wheeler--DeWitt equation. These orderings are in one-to-one correspondence with the Jacobians arising from field redefinitions of a set of canonical fields. For each operator ordering consistent with a path-integral measure, we identify a positive definite Hilbert-space inner product.
All such prescriptions define the same quantum theory, in the sense that they yield identical physical observables. We illustrate our formalism by applying it to de Sitter Jackiw--Teitelboim gravity and to the Starobinsky model.

		\end{quote}
		
		
	\end{titlepage}
	\newpage
	\setcounter{footnote}{0}
	\renewcommand{\thefootnote}{\arabic{footnote}}
	\setlength{\baselineskip}{0.7cm} \setlength{\parskip}{.2cm}
	
	\setcounter{section}{0}
	
	
	\section{Introduction}   
	
	In \Refe{WDW1D}, the issue of operator-ordering ambiguities \cite{DeWitt:1967yk,Christodoulakis:1984gp,Hawking:1985bk,Halliwell:1984eu,Nelson:2008vz,He:2015wla,PTV,Partouche:2021lyb,Partouche:2022kfi,Kehagias:2021wwr,Singh:2025ald,Mondal:2025qyd} in the formulation of the Wheeler--DeWitt (WDW) equation~\cite{DeWitt:1967yk} for the wavefunction of the Universe \cite{Vilenkin:1982de,Vilenkin:1983xq,Vilenkin:1984wp,Vilenkin:1987kf,Hartle:1983ai,Hawking:1984hk,Halliwell,Halli2,Hartle:2008ng,Turok1,Halli-Hartle,Linde:1990flp} has been studied in the simplest cases of minisuperspace models involving a single scale-factor degree of freedom~$a$. In the path-integral description of the quantum theory, the wavefunction of the Universe  can be written as
	\be
	\psi(\mathrm{a})=\int {\D N \over \Vol(\Diff)}\, \int_{\mathrm{a}_i}^{\mathrm{a}}\D a \; \,e^{{i\over \hbar}S[N,a,\dot{a}]}\, ,
	\label{psidef2}
	\ee
where $N$ is the lapse function, $S$ is the gravitational action, $\D a$ is a time-reparametrization invariant measure of the scale factor, $\Vol(\Diff)$ is  the volume of the time-repa\-ra\-me\-tri\-za\-tion group, and $\dot{a}$ is the time derivative of $a$. The Universe is taken to be closed, in order for the action to be finite. The ordering ambiguities can be seen to arise from the  infinite number of choices for the path-integral measure, 
which can be related to field redefinitions of the scale factor \cite{WDW1D,PTV,Partouche:2021lyb,Partouche:2022kfi,Kehagias:2021wwr,Halliwell}.  Indeed, given any invertible function $f$, we can define a new field variable $q=f(a)$ and construct the path-integral using the gauge invariant measure ${\cal{D}}q$.\footnote{Various methods can be applied to define the path-integral measures $\D a$ and $\D q$. Skeletonization is used in \Refs{WDW1D,Halliwell:1984eu}, while in \Refe{PTV} one expands the field fluctuations on bases of eigenstates of self-dual operators in order to integrate over the expansion coefficients. As shown in \Refe{WDW1D}, both methods yield identical results for the probability amplitudes and the physical observables.}  Despite the invariance of the classical action $S$ under such field redefinitions, the path-integral measure ${\cal{D}}q$ changes, since it is related to ${\cal{D}}a$ by a non-trivial Jacobian. As a result, we get a multitude of different quantum prescriptions leading to different wavefunctions for the Universe. These path-integral wavefunctions satisfy distinct WDW equations, which can be linked to different operator-ordering choices in constructing the quantum Hamiltonian \cite{WDW1D,PTV,Partouche:2021lyb,Partouche:2022kfi,Kehagias:2021wwr}. 

Despite receiving considerable attention \cite{DeWitt:1967yk,Christodoulakis:1984gp,Hawking:1985bk,Halliwell:1984eu,Nelson:2008vz,He:2015wla,PTV,Partouche:2021lyb,Partouche:2022kfi,Kehagias:2021wwr,Singh:2025ald,Mondal:2025qyd}, the problem of ordering ambiguities has yet to receive a complete resolution. To make progress in this direction, some of the authors showed in \Refe{WDW1D} that in the one-dimensional minisuperspace models, all path-integral prescriptions are physically equivalent.  This was achieved in several steps. Using  the flatness of the target space of the single degree of freedom, the WDW equation for each choice of the path-integral measure is determined exactly, to all orders in $\hbar$. To put it another way, all ordering ambiguities in the corresponding Hamiltonian are resolved. The Hilbert-space inner-product 
	\be
	\langle \psi_1|\psi_2\rangle = \int_{\mathcal{T}} \d \q \sqrt{-\gamma} \, \mu(\q)\, \psi_1( \q)^*\, \psi_2( \q)
	\ee
is then introduced, where $\gamma$ is the determinant of the metric of the target space $\mathcal{T}$ and $\mu(\q)$ is a real positive function derived exactly by imposing Hermiticity of the Hamiltonian. Finally, it is shown that the dressed wavefunction
	\be\label{dress}
	\Psi = \sqrt{\mu}\psi\, , 
	\ee
which can be used to write the inner product as
\be
	\langle \psi_1|\psi_2\rangle = \int_{\mathcal{T}} \d \q \sqrt{-\gamma} \,  \Psi_1( \q)^*\, \Psi_2( \q)\,,
	\ee
satisfies a universal WDW equation
	\be\label{univ}
	\boldsymbol{\mathcal{H}}\Psi \equiv -{\hbar^2 \over 2} \nabla^2 \Psi+ V\Psi=0\, .
	\ee
Indeed, the latter is free of any ambiguity associated with the path-integral measure or an operator ordering. In a nutshell,  while the wavefunctions $\psi_1$, $\psi_2$ and the inner-product function $\mu$ depend on the choice of path-integral measure, the dressed wavefunctions $\Psi_1$, $\Psi_2$ and the inner product itself turn out to be independent of this choice. Hence, all probability amplitudes and predictions for the physical observables are independent of the prescription. 

In this work, we show that the above universality  properties generalize to a larger class of minisuperspace models with $n\ge1$ degrees of freedom and a two-derivative kinetic term, for which the  target-space  $\T$ is flat.\footnote{By flat, we mean vanishing of the Riemann tensor of the minisuperspace manifold.} In the process, we clarify substantially the previous arguments of~\Refe{WDW1D}.  Despite the flatness restriction, this class includes interesting models, both from the theoretical and the phenomenological perspectives. 
To be specific, we consider minisuperspace models of gravity coupled to an arbitrary number of scalar fields and $p$-form field strengths, and allow space to be anisotropic. When the \mbox{$n$-dimensional} target space $\T$ is flat, there exist canonical fields $Q^I$, $I \in \{0,\dots,n-1\}$, such that the target-space metric is Minkowskian. In terms of an arbitrary field redefinition $q^i=F^i(\vec{Q})$, $i\in\{0,\dots, n-1\}$, the classical theory can be equivalently formulated in terms of a non-linear $\sigma$-model, with a one-dimensional base manifold parametrized by time. The Lagrangian reads
	\be
	L(N,\vec q,\dot{\vec q})=N\left({1\over2N^2}\,\gamma_{ij}(\vec q)\, \dot q^i\dot q^j-v(\vec q)\right) ,
	\ee
where $\gamma_{ij}$ is the new metric of the target space and $v(\vec q)$ is a potential.  However, as in the one-dimensional case, the wavefunction of the Universe can be defined using a multitude of choices for the  time-reparametrization invariant  path-integral measure, since the path-integral measures $\prod_{i=0}^{n-1}{\cal{D}}q^i$ and $\prod_{I=0}^{n-1}{\cal{D}}Q^I$ differ by a Jacobian.   Thanks to the flatness of $\T$, we show that for any choice of path-integral measure, the WDW equation for the wavefunction takes the exact following  form in $\hbar$:  
	\be
	{1\over J}\left[-{\hbar^2\over 2}\, \nabla^2(J\psi)+vJ\psi\right]=0\, ,\qquad \where\qquad 
	J=\left|\det {\partial F^i\over \partial \Q^I}\right|\, .
	\ee
Consistently, this equation is shown to correspond to a specific ordering of the quantum operators, when applying the canonical quantization on the classical Hamiltonian. Defining the Hilbert-space inner-product as 
 	\be
	\langle \psi_1|\psi_2\rangle = \int_\T \prod_{k=0}^{n-1}\d \q^k \sqrt{-\gamma} \, \mu(\vec \q)\, \psi_1(\vec \q)^*\, \psi_2(\vec \q)\,,
	\ee
the real positive function $\mu(\vec \q)$ is determined by imposing the Hermiticity of the Hamiltonian, which yields
\begin{equation}
		\mu =J^2\,.
	\end{equation}
Using the notations introduced earlier for the one-dimensional cases, the inner product can be  expressed solely in terms of the dressed wavefunctions~(\ref{dress}), which satisfy the universal WDW equation~(\ref{univ}). As a result, all physical observables and probability amplitudes are independent of the quantum prescription, to all orders in $\hbar$. Notice, furthermore, that in terms of the dressed wavefunctions, the inner product becomes the standard quantum mechanical inner product between wavefunctions.
	
	The plan of the paper is as follows. In~\Sect{model}, we describe the class of minisuperspace models we consider. They are based on $D$-dimensional Einstein gravity coupled to an arbitrary number of bosonic fields. We cast such models as non-linear $\sigma$-models involving a one-dimensional base manifold and a Lorentzian target space of dimension $n\ge 1$. We show that the path-integral formulation of the quantum theory leads to ambiguities, due to the infinite number of choices for the time-reparametrization invariant path-integral measure, which follow from field redefinitions. We proceed in~\Sect{WDWeq} to study models with a flat target space. For these models, we derive the WDW equation that the path-integral wavefunctions satisfy, irrespective of the choice of path-integral measure. The  result for the WDW equation is exact, valid to all orders in the $\hbar$ expansion. In~\Sect{equiv}, for any path-integral measure, we determine the Hilbert-space inner-product that insures the Hermiticity of the quantum Hamiltonian. We then use it to establish the equivalence of all quantum prescriptions. In~\Sect{appli}, we illustrate our results with inflationary models\mbox{---such} as the Starobinsky model~\cite{Starobinsky:1980te}---which all lead to a flat minisuperspace geometry.
We also apply our results to de Sitter Jackiw--Teitelboim (JT) gravity~\cite{Jackiw:1984je,Teitelboim:1983ux}, where the correct treatment of the factor-ordering ambiguities has so far remained an open problem. In~\Sect{compa}, we summarize our results, explain why they do not extend to non-flat target spaces, and compare them with previous results in the literature.

	
	\section{Minisuperspace models}
	\label{model}

Our starting point is Einstein gravity in $D$ dimensions, coupled exclusively to bosonic degrees of freedom, for the sake of simplicity. Setting $8\pi G_{\rm N}=1$, where $G_{\rm N}$ is Newton's constant, the action takes the generic form\footnote{In \Sect{JTex}, we will apply our results to the de Sitter JT gravity, which involves an alternative form of action.}
\be\begin{split}
S=\int_{\M}\d^D x\,\sqrt{-g}\,\bigg(&{R\over 2}-{1\over 2} \, G_{AB}(\vec{\mathit{\Phi}})\, \partial_\mu\mathit{\Phi}^A\partial^\mu\mathit{\Phi}^B-\V(\vec{\mathit{\Phi}})\\
&+\sum_\alpha{1\over 2(p_\alpha+1)!}\, f^{(\alpha)}(\vec{\mathit{\Phi}})\, F^{(\alpha)}_{\mu_0\dots\mu_{p_\alpha}}F^{(\alpha)\mu_0\dots\mu_{p_\alpha}}\bigg)
+\int_{\partial\mathcal{M}}\d^3y \,\sqrt{h}\,K\, , 
\end{split}
\label{SE}
\ee
where $\mathcal{M}$ is a spatially-closed Lorentizian manifold of coordinates $x^\mu$, metric $g_{\mu\nu}$ and Ricci scalar $R$, while $\partial\mathcal{M}$ is a spacelike boundary of coordinates $y^m$, induced metric $h_{mn}$ and trace of the extrinsic curvature $K$. The action includes scalar fields $\mathit{\Phi}^A$, with $\sigma$-model metric $G_{AB}(\vec{\mathit{\Phi}})$ and potential $\V(\vec{\mathit{\Phi}})$, along with field strengths $F^{(\alpha)}_{\mu_0\dots\mu_{p_\alpha}}$ of $p$-form gauge potentials $A^{(\alpha)}_{\mu_1,\dots,\mu_{p_\alpha}}$, with couplings $f^{(\alpha)}(\vec{\mathit{\Phi}})$.

In this work, we focus on a minisuperspace version of this theory, in which the Universe is restricted to be homogeneous,\footnote{More precisely, it is finiteness of the Universe that is required, to ensure that the action remains finite, so that the path-integral wavefunction is well defined.} and the spacetime metric, scalar fields, and gauge potentials are assumed to take the following forms
	\be\begin{split}
	&\d s^2=-N(x^0)^2\, {\d x^0}^2+g_{ab}(x^0)\, \d x^a\d x^b\, ,\\
	&\mathit{\Phi}^A(x^0)\, , \qquad A^{(\alpha)}_0=0\, , \qquad A^{(\alpha)}_a(x^0)\, ,
	\end{split}
	\ee 
	where $N$ is the lapse function and $a,b\in\{1,\dots,D-1\}$ are spatial indices. Integrating by parts, all boundary terms occurring in the action cancel each other, and the Lagrangian involves only first derivatives. For notational simplicity, we collectively relabel the components $g_{ab}$, the scalars $\mathit{\Phi}^A$ and the components $A^{(\alpha)}_a$ as fields $q^i(x^0)$, $i\in\{0,\dots,n-1\}$. In this case, the action can be recast in the form of a non-linear $\sigma$-model 
	\begin{align}
	S&=\int \d x^0\, L\, , \\
	\label{sigma}\where\qquad L(N,\vec q,\dot{\vec q})&=N\left({1\over2N^2}\,\gamma_{ij}(\vec q)\, \dot q^i\dot q^j-v(\vec q)\right) . 
	\end{align}
	In this expression, $\gamma_{ij}$ is the metric of the target space $\T$ spanned by the fields $q^i$ (\ie the minisuperspace), $v$ is a scalar potential and the dots denote time derivatives. Notice that $\T$ is  Lorentzian~\cite{DeWitt:1967yk} and has a boundary. This is easily seen for instance in the simplest model, for which $n=1$, corresponding to a homogeneous and isotropic Universe involving only a scale-factor $a(x^0)$, with no other degree of freedom. Indeed, the kinetic term of the scale factor has an opposite sign, as  compared to that of a scalar field, and the scale factor satisfies $a\ge 0$, which implies that the target space $\T$ reduces to a ray. In this example where $n=1$, the potential involves spatial curvature and cosmological constant terms~\cite{DeWitt:1967yk,WDW1D}.
	
	The quantum version of the generic model with Lagrangian~(\ref{sigma}) can be formulated in terms of the wavefunction of the Universe~\cite{Vilenkin:1982de,Vilenkin:1983xq,Vilenkin:1984wp,Vilenkin:1987kf,Hartle:1983ai}. The latter can be defined as a Lorentzian path-integral, which is a sum over all trajectories of the lapse function $N(x^0)$ and fields $\vec q(x^0)$,  defined on an arbitrary interval $[x^0_\i,x^0_\f]$ and satisfying the boundary conditions
	\be
	\vec q(x^0_\i)=\vec\q_\i\, , \qquad \vec q(x^0_\f)=\vec\q\, .
	\ee
	In the above equation, $\vec\q_{\rm i}$, $\vec\q$ are arbitrary points in the manifold $\T$. In our notations, and everywhere in this work, for a given metric $\gamma_{ij}$, we denote fields in the target space $\T$ by {\it italic} fonts such as $\vec q$, and points of the manifold $\T$ by {\rm roman} fonts such as $\vec \q$. Using these conventions, the wavefunction can be written as
	\be
	\psi(\vec \q)=\int {\D N \over \Vol(\Diff)}\, \int_{\vec q(x^0_\i)=\vec \q_\i}^{\vec q(x^0_\f)=\vec \q}\;\, \prod_{k=0}^{n-1}\D q^k \; \,e^{{i\over \hbar}\int_{x^0_\i}^{x^0_\f}\d x^0\, L(N,\vec q,\dot {\vec q})}\, ,
	\label{psidef}
	\ee
	where $\Vol(\Diff)$ is the volume of the diffeomorphism group associated with the line segment of time $[x^0_\i,x^0_\f]$. In this expression, $\vec \q_\i$ characterizes the state of the Universe. Notice that the paths of the lapse function $N(x^0)$ can be grouped in equivalence classes labeled by the length $\ell\in\R_+$,
	\be
	\ell=\int_{x^0_\i}^{x^0_\f}\d x^0 N\, , 
	\ee
	which is invariant under diffeomorphisms. As a result, the path-integral of the lapse function, divided by $\Vol(\Diff)$, can be replaced by an integral over the equivalence classes~$\ell$. In practice, we first choose a gauge, \ie a representative of each class, which can be
	\be
	N(x^0)\equiv \ell\, , \quad \mbox{defined on}\quad [x^0_\i, x^0_\f]=[0,1]\, .
	\ee
	We then apply the method of Faddeev and Popov, as described in detail in \Refe{PTV}. This amounts to computing a  Faddeev-Popov determinant, which turns out to be trivial, \mbox{$\Delta_{\rm FP}(\ell)=1$}. As a result, the wavefunction takes the form
	\be
	\psi(\vec \q)=\int_0^{+\infty}\d\ell  \, \int_{\vec q(0)=\vec \q_\i}^{\vec q(1)=\vec \q}\;\, \prod_{k=0}^{n-1}\D q^k \; e^{{i\over \hbar}\int_{0}^{1}\d x^0\, L(\ell,\vec q,\dot {\vec q})}\, .
	\label{psi2}
	\ee
We have chosen to formulate the wavefunction in terms of a Lorentzian path integral, in order to avoid problems and difficulties with the unboundedness of the Euclidean gravity action~\cite{Turok1,Halli-Hartle}. Due to the oscillatory nature of the phase $e^{iS/\hbar}$, the Lorentzian path integral can be conditionally convergent (at least in some simple cases). Using Picard--Lefschetz theory, the path of integration with respect to $\ell$ can be deformed from the real positive axis to a curve in the complex plane, in order to produce an absolutely convergent integral~\cite{Turok1}. 
	
	The above definition of the wavefunction is however ambiguous, due to the infinite number of possible choices of path-integral measure.  To make this point explicit, let us perform an invertible  field redefinition
	\be
	q^i=f^i(\vec {\tilde q}\,), \qquad i\in\{0,\dots,n-1\}\,.
	\ee
	Under such a transformation, the  classical action is invariant, while the path-integral measure satisfies
	\be
	\prod_{k=0}^{n-1}\D q^k = \prod_{k=0}^{n-1}\D \tilde q^k  \times \J  ,
	\ee
	where $\J$ is a Jacobian, which can be formally written as 
	\be
	\J =\prod_{x^0\in[0,1]}\left|\det {\partial f^i\over \partial \tilde q^j}\big(\vec{\tilde q}(x^0)\!\big)\right|\, .
	\ee 
	Therefore, the wavefunction in \Eq{psi2} can be expressed as
	\be
	\psi\big(\vec \q(\vec {\tilde \q}\,)\!\!\:\big)=\int_0^{+\infty}\d\ell  \, \int_{\vec {\tilde q}(0)=\vec {\tilde \q}_\i}^{\vec {\tilde q}(1)=\vec {\tilde \q}}\;\, \prod_{k=0}^{n-1}\D \tilde q^k \, \J\; \,e^{{i\over \hbar}\int_{0}^{1}\d x^0\, \tilde L(\ell,\vec {\tilde q},\dot {\vec {\tilde q}})}\, ,
	\label{psi3}
	\ee
	where $\tilde L$ is the Lagrangian involving the transformed metric $\tilde \gamma_{ij}$, 
	 \be
		\tilde L\big(\ell,\vec {\tilde q},\dot{\vec {\tilde q}}\,\big)=\ell\left({1\over2\ell^2}\,\tilde \gamma_{ij}(\vec {\tilde q}\,)\, \dot {\tilde q}^i\dot {\tilde q}^j-v\big(\vec f(\vec {\tilde q}\,)\!\!\:\big)\right) ,
	\ee
	and where the boundary conditions of the new fields $\vec{\tilde q}$ satisfy
	\be
	\vec \q=\vec f(\vec {\tilde \q})\, , \qquad \vec \q_\i=\vec f(\vec {\tilde \q}_\i)\, .
	\label{f}
	\ee
	Since the fields $\vec q$ and $\vec{\tilde q}$ are on equal footing, instead of using \Eq{psi2}, one could have chosen the alternative definition of wavefunction
	\be
	\tilde \psi(\vec {\tilde \q})=\int_0^{+\infty}\d\ell  \, \int_{\vec {\tilde q}(0)=\vec {\tilde \q}_\i}^{\vec {\tilde q}(1)=\vec {\tilde \q}}\;\, \prod_{k=0}^{n-1}\D \tilde q^k \; e^{{i\over \hbar}\int_{0}^{1}\d x^0\, \tilde L(\ell,\vec {\tilde q},\dot {\vec {\tilde q}})}\, ,
	\ee
	which is exactly \Eq{psi3} {\it without} the Jacobian $\J$. As a result, $\psi\big(\vec \q(\vec {\tilde \q}\,)\!\big)$ and $\tilde \psi(\vec {\tilde \q})$ are not equal and it is not clear whether the quantum theories that result from these different choices of path-integral measures are equivalent or not. The next two sections will provide an answer to this question, in the case where the target space $\T$ is flat. 
	
	
	\section{Wheeler--DeWitt equation}
	\label{WDWeq}
	
	We have seen in the previous section that the wavefunction depends on the choice of path-integral measure of the degrees of freedom. Therefore, for each prescription, the wavefunction of the Universe is expected to be the solution of a distinct WDW  equation. In the present section, we determine this differential equation.
	
	To reach this goal, we find convenient to work in cosmological time defined as
	\be
	t=\ell x^0\, , 
	\ee
	and first derive an equation for the amplitude
	\be
	u(\vec \q_\i,\vec \q,\ell)=\int_{\vec q(0)=\vec \q_\i}^{\vec q(1)=\vec \q}\;\, \prod_{k=0}^{n-1}\D q^k \; e^{{i\over \hbar}\int_{0}^
		\ell\d t\,   \left({1\over 2}\gamma_{ij}(\vec q){\d q^i\over \d t}{\d q^j\over \d t}-v(\vec q) \scriptstyle \right)} .
	\label{u}
	\ee
For reasons that will be explained in the concluding section, {\it let us restrict for now on $\sigma$-models where the target space $\T$ is flat.} Although the class of theories satisfying this constraint is very restrictive, it nevertheless contains  interesting models, as seen in~\Sect{appli}. When the manifold $\T$ is flat, a change of coordinate brings the metric into a Minkowski form, $\eta_{IJ}$, \mbox{$I,J\in\{0,\dots,n-1\}$}. From the point of view of the degrees of freedom, this corresponds to a field redefinition 
	\be
	\label{toflat}
	q^i=F^i(\vec Q\,), \qquad i\in\{0,\dots,n-1\}\,,
	\ee
	where the new fields $Q^I$, $I\in\{0,\dots,n-1\}$, are canonical. An alternative form of the amplitude is thus
	\be
	\begin{split}
		u(\vec \q_{\i},\vec \q,\ell)&=\int_{\vec Q(0)=\vec \Q_\i}^{\vec Q(1)=\vec \Q}\;\, \prod_{K=0}^{n-1}\D Q^K \; \prod_{t\in[0,\ell]}J\big(\vec Q(t)\big)\;\, e^{{i\over \hbar}\int_{0}^
			\ell\d t\,  \left({1\over 2}\eta_{IJ}{\d Q^I\over \d t}{\d Q^J\over \d t}-V(\vec Q) \scriptstyle \right)}\\
		&\equiv U(\vec \Q_{\i},\vec \Q,\ell)\, ,
	\end{split}
	\label{U}
	\ee
	where the Roman type quantities and potential are defined as
	\be
	\vec \q=\vec F(\vec \Q)\, , \qquad \vec \q_\i=\vec F(\vec \Q_\i)\, ,\qquad V(\vec Q)=v\big(\vec F(\vec Q)\big)\, ,
	\label{F}
	\ee
	while the Jacobian is written in terms of the function\footnote{The function 
		$J$ should not be confused with the index $J$ of the flat coordinates and canonical fields.}
	\be
	J(\vec \Q)=\left|\det {\partial F^i\over \partial \Q^I}(\vec \Q)\right|\, .
	\label{J(Q)}
	\ee
	
	The differential equation satisfied by $U(\vec \Q_{\i},\vec \Q,\ell)$ can be derived by discretizing the interval of cosmological time $[0,\ell]$ into $A\ge 1$ parts of duration $\varepsilon$. Denoting $\vec Q_\alpha$, $\alpha\in\{0,\dots,A\}$, the values of the fields $\vec Q$ at $t=\alpha \varepsilon$, a discretized version of the amplitude is
	\be
	\label{disc}
	\begin{split}
		U^{(A)}(\vec \Q_\i,\vec \Q,\ell)&=\prod_{\alpha'=1}^{A-1} \left(\int_\T {\prod_{K=0}^{n-1}\d Q_{\alpha'}^K\over \N(\varepsilon,\vec \Q_\i,\vec \Q)} \, J(\vec Q_{\alpha'})\right)\espD\\
		&~~~~~~~~~~~~~~~~~~~~~~~\times e^{{\textstyle{i\over \hbar}}{\sum_{\alpha=0}^{A-1}}\varepsilon \big[{1\over 2}\,\eta_{IJ}{Q^I_{\alpha+1}-Q^I_{\alpha}\over \varepsilon}{Q^J_{\alpha+1}-Q^J_{\alpha}\over \varepsilon}-V{\textstyle(}{\vec Q_{\alpha}+\vec Q_{\alpha+1}\over 2}\textstyle)\big]}  \, ,\esp\\
		\where \qquad \varepsilon&={\ell\over A} \, , \qquad \vec Q_0=\vec \Q_\i\, , \qquad \vec Q_A=\vec \Q\, .
	\end{split}
	\ee 
	Note that  for later regularization of the path-integral measure, a normalization factor $\N$ is introduced for each variable $\vec Q_{\alpha'}$ on which we integrate. It turns out to be useful to separate the last slice of time, in order to relate the amplitudes for $A$ and $A-1$ as follows:  
	\be
	\begin{split}
		U^{(A)}(\vec \Q_\i,\vec \Q,\ell)&=\int_\T {\prod_{K=0}^{n-1}\d Q_{A-1}^K\over \N(\varepsilon,\vec \Q_\i,\vec \Q)} \, J(\vec Q_{A-1})\; e^{{i\over 2 \hbar \varepsilon} \eta_{IJ}(Q^I_{A-1}-\Q^I)(Q^J_{A-1}-\Q^J)}\; e^{-{i\over \hbar}\varepsilon V{\textstyle(}{\vec Q_{A-1}+\vec \Q\over 2}\textstyle)}  \\
		&\hspace{7cm}\times U^{(A-1)}(\vec \Q_\i,\vec Q_{A-1},\ell-\varepsilon)\, .
	\end{split}
	\label{eq}
	\ee 
	When $\varepsilon$ is small, fast oscillations of the first exponential in the integrand imply that the only contributions to the integral arise for $|Q^I_{A-1}-\Q^I|\lesssim \sqrt{2\hbar\varepsilon}$. Therefore, the second exponential  can be safely approximated by a double expansion in $\varepsilon$ and \mbox{$Q^I_{A-1}-\Q^I$},
	\be\label{ea}
	e^{-{i\over \hbar}\varepsilon V{\textstyle(}{\vec Q_{A-1}+\vec \Q\over 2}\textstyle)}  =1-{i\over \hbar}\,\varepsilon\left[V(\Q)+\O(\vec Q_{A-1}-\vec \Q)\right]\!+\O(\varepsilon^2)\, ,
	\ee
	while it is enough to write for the remaining factors in the integrand 
	\be\label{eb}
	\begin{split}
		&J(\vec Q_{A-1})\, U^{(A-1)}(\vec \Q_\i,\vec Q_{A-1},\ell-\epsilon)=\\
		&\!\!\left[1+(Q_{A-1}^I-\Q^I)\, {\partial\over \partial Q^I_{A-1}}+{1\over 2}\,(Q^I_{A-1}-\Q^I)(Q^J_{A-1}-\Q^J)\, {\partial^2\over \partial Q^I_{A-1}\partial Q^J_{A-1}}+\O\big((\vec Q_{A-1}-\vec \Q)^3\big)\!\right]\esp\\
		&\hspace{8.1cm}\times \big[J(\vec Q_{A-1})\, U^{A-1}(\vec\Q_\i,\vec Q_{A-1},\ell-\epsilon)\big]\Big|_{\vec Q_{A-1}=\vec\Q}\, .\esp
	\end{split}
	\ee
	We are going to integrate term by term the right-hand-side of \Eq{eq} by applying the Gaussian formulas
	\begin{align}
		\int_{\R^{1,n-1}} \Big(\prod_{K=0}^{n-1}\d Q^K\Big) \; e^{-c \,\eta_{IJ}Q^IQ^J}&=\Big({\pi\over c}\Big)^{n\over 2}\sqrt{1\over \det \eta}\, , \\
		\int_{\R^{1,n-1}} \Big(\prod_{K=0}^{n-1}\d Q^K\Big) \; Q^LQ^M\; e^{-c\, \eta_{IJ}Q^IQ^J}&={\eta^{LM}\over 2c}\Big({\pi\over c}\Big)^{n\over 2}\sqrt{1\over \det \eta}\, .\esp
	\end{align}
	One may worry about the fact that since the flat manifold $\T$ has a boundary, it is only made of subregions of $\R^{1,n-1}$, with possible identifications like in orbifold constructions. However, as we are interested in the $\varepsilon \to 0$ limit, replacing $\T$ by $\R^{1,n-1}$ only introduces errors that are exponentially small in $\varepsilon$. Moreover, the use of the Gaussian formulas requires a regularization scheme to insure convergence of the integrals. This is done by replacing, for each pair $(I,J)$,
	\be
	{i\over 2 \hbar \varepsilon} \,\eta_{IJ}\longrightarrow{i\over 2 \hbar \varepsilon} \,\eta_{IJ}\, (1+i\kappa)\, ,\qquad \where\qquad \kappa>0\, ,  
	\ee
	and then taking $\kappa\to 0$ after integration. By proceeding in this way, \Eq{eq} leads to
	\be
	\begin{split}
		U^{(A)}(\vec \Q_\i,\vec \Q,\ell)=&\, {1\over \N(\varepsilon,\vec \Q_\i,\vec \Q)}\,\Big({2\pi\hbar\varepsilon\over -i}\Big)^{n\over 2}\sqrt{1\over \det \eta}\\
		&~~~~~\times \Big(1-i\,{\varepsilon\over \hbar}\, V(\vec \Q)-{\hbar\varepsilon\over 2i}\,\eta^{IJ}{\partial\over \partial Q^I_{A-1}\partial Q^J_{A-1}}+{\cal{O}}(\varepsilon^2)\Big)\\
		&~~~~~\times\big[J(\vec Q_{A-1})\, U^{(A-1)}(\vec \Q_\i,\vec Q_{A-1},\ell-\varepsilon)\big]\Big|_{\vec Q_{A-1}=\vec \Q}\, .
	\end{split}
	\label{eq1}
	\ee
	In the large $A$ limit (small $\varepsilon$ limit), this expression imposes a constraint on the normalization function $\N$,
	\be
	U(\vec \Q_\i,\vec \Q,\ell)\underset{\varepsilon\to 0}\sim{1\over \N(\varepsilon,\vec \Q_\i,\vec \Q)}\,\Big({2\pi\hbar\varepsilon\over -i}\Big)^{n\over 2}\sqrt{1\over \det \eta}\; J(\vec \Q)\; U(\vec \Q_\i,\vec \Q,\ell)\, .
	\ee
	Making the allowed choice 
	\be
	\N(\varepsilon,\vec \Q_\i,\vec \Q) = \Big({2\pi\hbar\varepsilon\over -i}\Big)^{n\over 2}\sqrt{1\over \det \eta}\; J(\vec \Q)\, ,
	\ee
	\Eq{eq1} yields
	\be
	\begin{split}
		&i\hbar\,{U^{A-1}(\vec \Q_\i,\vec\Q,\ell-\varepsilon)-U^{(A)}(\vec\Q_\i,\vec \Q,\ell)\over -\varepsilon}=\\
		&\,~~~~~~~~~~~~~~~~~~~~-{\hbar^2\over 2}\, {\eta^{IJ}\over J(\vec\Q)}\, {\partial^2\over \partial Q^I_{A-1}\partial Q^J_{A-1}}\big[J(\vec Q_{A-1})\, U^{(A-1)}(\vec \Q_\i,\vec Q_{A-1},\ell)\big]\Big|_{\vec \Q_{A-1}=\Q}\esp\\
		&\,~~~~~~~~~~~~~~~~~~~~+V(\vec \Q)\, U^{(A-1)}(\vec\Q_\i,\vec \Q,\ell)+\cal O(\varepsilon)\, .\esp
	\end{split}
	\ee
	We are now ready to take the continuum limit $A\to +\infty$ ($\varepsilon\to 0$ limit), to obtain the differential equation for $U(\vec\Q_\i,\vec \Q,\ell)$, 
	\be
	i\hbar\,{\partial U\over \partial\ell}= {1\over J}\left[-{\hbar^2\over 2}\,  \eta^{IJ} {\partial^2(J U)\over \partial \Q^I\partial\Q^J}+VJU\right].
	\label{eqU}
	\ee
	For reasons that will be explained in \Sect{equiv}, even if it looks natural, we choose not to multiply this equation by $J(\vec \Q)$. In the right-hand side, we recognize a Dalembertian in flat coordinates $\vec \Q$.  As a result, the equation for $u(\vec \q_\i,\vec \q,\ell)$ is
	\be
	i\hbar\,{\partial u\over \partial\ell}= {1\over J}\left[-{\hbar^2\over 2}\, \nabla^2(J u)+vJu\right],
	\label{equ}
	\ee
	where $\nabla$ is the Levi-Civita connection associated with the metric $\gamma_{ij}$ and $J$ stands for $J\big(\vec\Q(\vec \q)\big)$. 
	
	We next derive the WDW equation for the wavefunction in \Eq{psi2}, 
	\be
	\psi(\vec \q) = \int_0^{+\infty}\d\ell \;u(\vec \q_\i,\vec \q,\ell)\, .
	\ee
	Integrating \Eq{equ} over the lapse of cosmological time $\ell$, we  obtain
	\be
	i\hbar\, \big[u(\vec \q_{\i},\vec \q,\ell)\big]^{\ell=+\infty}_{\ell=0}= {1\over J}\left[-{\hbar^2\over 2}\, \nabla^2(J \psi)+vJ\psi\right],
	\ee
	where the left-hand side requires a regularization scheme. To define the value of $u$ at $\ell=+\infty$, we use a formal expression for the path-integral in $u$, we add a small imaginary part to $\ell$, and we consider a double limit. In practice, we write
	\be
	\begin{split}
		u(\vec \q_\i,\vec \q,+\infty)&=\lim_{\kappa\to 0^+}\lim_{\ell\to +\infty}\prod_{x^0\in(0,1)}\int_\T\prod_{k=0}^{n-1}\d q^k(x^0)   \; e^{{i\over \hbar} \scriptstyle \big[{1\over 2\ell}\gamma_{ij}\dot q^i\dot q^j-\ell(1-is\kappa\scriptstyle ) v\scriptstyle \big]\!\big|_{x^0}} ,\\
		\where \qquad s(\vec q)&\equiv \sign\!\big(v(\vec q)\big)\, , \qquad \vec q(0)=\vec \q_\i\, ,  \qquad\vec q(1)=\q\, .\esps
	\end{split}
	\ee
	Due to the extra factor $e^{-{\kappa\over \hbar}\ell |v(\vec q)|}$ in the integrand, we find that  $u(\vec \q_\i,\vec \q,+\infty)=0$. To define the value of $u$ at $\ell=0$, we approach the origin along the imaginary axis, as follows:
	\be
	\begin{split}
		u(\vec \q_\i,\vec \q,0)&=\lim_{\kappa\to 0^+}\prod_{x^0\in(0,1)}\int_\T\prod_{k=0}^{n-1}\d q^k(x^0)   \; e^{{i\over \hbar} \scriptstyle \big[{1\over 2is\kappa}\gamma_{ij}\dot q^i\dot q^j -is\kappa v\scriptstyle \big]\!\big|_{x^0}} ,\\
		\where \qquad s(\vec q, \dot{\vec q})&\equiv -\sign\!\big(\gamma_{ij}\dot q^i\dot q^j\big)\, .\esps
	\end{split}
	\ee
	Due to the factor $e^{-{1\over 2\hbar\kappa}|\gamma_{ij}\dot q^i\dot q^j|}$ in the integrand, we find that  $u(\vec \q_\i,\vec \q,+\infty)=0$. As a result, the wavefunction $\psi(\vec\q)$ satisfies the WDW equation
	\be
	{\cal H}\psi\equiv{1\over J}\left[-{\hbar^2\over 2}\, \nabla^2(J\psi)+vJ\psi\right]=0\, ,
	\label{wdwpsi}
	\ee
	which can be alternatively written as 
	\be
	{\cal H}\psi\equiv\left\{\!-{\hbar^2\over 2}\left[\nabla^2+{2\over J}\, \nabla^i J\, \nabla_i+{1\over J}\,\nabla^2 J \right]+ v\right\} \psi=0\, .
	\label{wdwpsi2}
	\ee
	
	As anticipated, since the wavefunction $\psi(\vec \q)$ defined in \Eqs{psidef} or~(\ref{psi2}) depends on the choice of path-integral measure appearing in \Eq{U},
	\be
	\prod_{k=0}^{n-1}\D q^k\equiv \prod_{K=0}^{n-1}\D Q^K \; \prod_{t\in[0,\ell]}J\big(\vec Q(t)\big)\, ,
	\label{measure}
	\ee
	we  find that the WDW equation depends on the Jacobian  $J$. Some remarks are in order:
	
	\begin{itemize}
		\item Beside the universal terms $-(\hbar^2/2)\nabla^2\psi$ and $v\psi$ present in \Eq{wdwpsi2}, there are two $J$-dependent contributions. To estimate their effects, one can use the WKB ansatz 
		\be\
		\psi(\vec \q)=e^{{i\over\hbar}\left(f_0(\vec\q)+{\hbar\over i}f_1(\vec \q)+\O(\hbar^2)\right)}
		\ee
		in the equation. This leads to a differential condition for $f_1$ (the 1st-order term in $\hbar$) that involves the term $-(\hbar^2/2)(2/ J)\, \nabla^i J\, \nabla_i\psi$. Hence, the latter already affects the wavefunction at the semiclassical level. On the contrary, the extra term $-(\hbar^2/2)\nabla^2J/J$ in the equation is a 2d-order correction in $\hbar$ to the potential $v$. Therefore, it affects the wavefunction only at this order and beyond.
		
		\item Consistently, the function $J$ cannot be gauged away by using the reparametrization symmetry of the equation. Indeed, using an arbitrary coordinate system $\vec{\hat \q}$ of the target space $\T$, the equation becomes
		\be
		{\cal H}\psi\equiv\left\{\!-{\hbar^2\over 2}\left[\hat \nabla^2+{2\over J}\, \hat \nabla^i J\, \hat \nabla_i+{1\over J}\,\hat\nabla^2 J \right]+v\right\} \psi=0\, ,
		\label{wdwhat}
		\ee
		where $\hat \nabla$ is the Levi-Civita connection corresponding to the transformed metric $\hat \gamma_{ij}$, while $J$, $v$ and $\psi$ stand for $J\big(\vec\Q(\vec{\hat\q})\big)$, $v\big(\vec \q(\vec{\hat\q})\big)$  and  $\psi\big(\vec \q(\vec {\hat \q})\!\!\:\big)$. We insist on the fact that {\it while the equation is covariant with respect to a change of coordinates, it is not  ``covariant'' with respect to a change of path-integral measure $\prod_{k=0}^{n-1}\D q^k\to \prod_{k=0}^{n-1}\D \tilde q^k$, since the associated wavefunctions are distinct, as explained  at the end of \Sect{model}.}
		
	\end{itemize}
	
	
	\section{Equivalence of the quantum theories}
	\label{equiv}
	
	We have seen in the previous sections that for a given path-integral measure $\prod_{k=0}^{n-1}\D q^k$, which is characterized by the Jacobian $J$ defined in \Eq{J(Q)}, the WDW equation and the wavefunction depend on $J$. It is therefore legitimate to ask whether different choices of functions $J$ yield distinct quantum theories, even if they are  associated with the same classical model. To answer this question, we ask whether the probability amplitudes depend on $J$ or not. 
	
	The probability amplitude between an initial and a final state is the inner product of the associated wavefunctions $\psi_1(\vec \q)$, $\psi_2(\vec \q)$. More precisely, the inner product $\langle \psi_1|\psi_2\rangle $ can be interpreted as the probability amplitude to find the Universe in state $|\psi_1\rangle $, if initially the Universe is in state $|\psi_2\rangle$. The probability vanishes when the states are orthogonal and the inner product vanishes.  Since the wavefunctions depend on the choice of Jacobian $J$, the definition of the inner product of their Hilbert space may also depend on $J$. Therefore, we consider an ansatz for the inner product 
	\be
	\label{inner}
	\langle \psi_1|\psi_2\rangle = \int_\T \prod_{k=0}^{n-1}\d \q^k \sqrt{-\gamma} \, \mu(\vec \q)\, \psi_1(\vec \q)^*\, \psi_2(\vec \q)\,,
	\ee
	where $\prod_{k=0}^{n-1}\d \q^k \sqrt{-\gamma} \, \mu(\vec \q)$ is a measure involving a function~$\mu$, which depends \apriori on~$J$. Since the inner product is positive definite, $\mu$ must be real positive. In the following, we first determine $\mu$  by imposing that the inner product is consistent with the Hermiticity of the quantum Hamiltonian.\footnote{In the literature, one often employs the Klein-Gordon norm~\cite{DeWitt:1967yk} in order to reformulate a notion of conservation of probability in terms of a Klein--Gordon equation on the superspace, where ``time'' is replaced by an internal variable (\eg a scalar field). However, the Klein-Gordon norm is not positive definite and this approach causes problems.}
	
	To this end, we need to identify the quantum Hamiltonian. It is well known that its representation acting on the wavefunctions of the Hilbert space is nothing but the WDW equation~\cite{DeWitt:1967yk}. 
	However, since the WDW equation~(\ref{wdwpsi}) involves the function $J$, it is interesting to see how $J$ emerges by lifting the ambiguities arising in the canonical quantization of the classical Hamiltonian. For the classical Lagrangian defined in \Eq{sigma}, the  momenta conjugate to the fields $q^i$, $i\in\{0,\dots,n-1\}$, are 
	\be
	\pi_i = {\partial L\over \partial \dot q^i}=\frac{1}{N}\, \gamma_{ij}\dot{q}^j\, ,
	\ee
	and the classical Hamiltonian satisfies
	\be
	\begin{split}
		H&=\dot q^i\pi_i-L\\
		&={1\over 2N}\, \gamma_{ij} \dot q^i\dot q^j+Nv\\
		&=-N\, {\partial L\over \partial N}\, .
	\end{split}
	\ee
	Due to the last equality, the on-shell classical Hamiltonian is constrained to vanish, 
	\be
	{H\over N} ={1\over2}\,\gamma^{ij}(\vec q)\, \pi_i\pi_j+v(\vec q) = 0\, .
	\ee
	Note that since the classical fields $q^i$ and the conjugate momenta $\pi_j$ commute, there exist an infinite number of equivalent ways to write the classical Hamiltonian. However, upon canonical quantization, all these ways give rise to distinct Hamiltonian constraints, since conjugate variables cease to commute. In order to see that \Eq{wdwpsi} arises for a specific choice of ordering,  we use the fact that $(\partial Q^I/\partial q^j)(\partial q^k/\partial Q^I)=\delta^k_j$ (see \Eq{toflat}) to write
	\be
	\label{lift am}
	\gamma^{ij}\pi_i\pi_j = {1\over J} \, \gamma^{ij}\;{\partial Q^I \over \partial q^j}\; \pi_i\; {\partial q^k\over \partial Q^I}\;\pi_k \, J\, .
	\ee
	Upon canonical quantization, which amounts to replacing
	\be
	q^i \longrightarrow \q^i\, , \qquad \pi_i\longrightarrow -i\hbar {\partial\over \partial \q^i}\, , 
	\ee
	the right-hand side of \Eq{lift am} leads to the Hamiltonian constraint on the wavefunctions
	\be
	-{\hbar^2\over 2}\, {1\over J}\, \gamma^{ij}\, {\partial \Q^I \over \partial \q^j}\,{\partial\over \partial \q^i}\!\left[{\partial \q^k\over \partial \Q^I}\, {\partial(J\psi)\over \partial \q^k}\right]+v\psi = 0\, .
	\ee
	By noticing that $\gamma^{ij}=\eta^{KL}(\partial \q^i/\partial \Q^K)(\partial \q^j/\partial \Q^L)$, the above formula becomes 
	\be
	-{\hbar^2\over 2}\, {1\over J}\, \eta^{KI} {\partial^2(J\psi)\over \partial \Q^K\partial\Q^I}+V\psi = 0\, ,
	\ee
	which is the WDW equation~(\ref{wdwpsi}) written in flat coordinates. The left-hand side is thus the representation of the quantum Hamiltonian acting on the wavefunctions as written in \Eq{wdwpsi}, or in any coordinate system as in  \Eq{wdwhat}. Note that had we multiplied \Eq{wdwpsi} by $J$, as envisaged below \Eq{eqU}, the left-hand side would not have been the Hamiltonian.
	
	Before imposing Hermiticity of the Hamiltonian to determine the function $\mu$, one may ask why we do not impose Hermiticity of simpler operators, such as the position operator or its conjugate, which are represented by $\q^i$ and $-i\hbar \,{\partial/\partial \q^i}$. The reason is that {\it these operators are not endomorphisms of the Hilbert space.} To understand why, recall that the wavefuntions are all solutions of the WDW equation. However, given a wavefunction $\psi(\vec\q)$, the functions $\q^i\psi(\vec\q)$ and $-i\hbar\,{\partial\psi/ \partial\q^i}$ are not solutions of the equation, unless $\psi$ is the null function. Hence, the position operator and its conjugate map wavefunctions outside the Hilbert space, and imposing Hermiticity to them, with respect to the Hilbert space inner product, does not make sense. On the contrary, the action of the Hamiltonian on any wavefunction leads to the null function, ${\cal H}\psi=0$, which is a solution of the WDW equation (even though it is not normalizable). This is an observable property provided it is Hermitian.\footnote{ Note that $q^i$ and $p^i$ are not well-defined as quantum mechanical operators when there exists a constraint of the type $q^i\geq 0$ (see \eg \Refe{Capri}), and some care must be taken to properly define them. This issue has appeared in the recent literature, \eg in~\Refe{Chandrasekaran:2022cip}, where it was argued that the naive definition of $p$ and $q$ is a good approximation for states that are supported mostly away from $q^i=0$. In our setup, such an issue appears when one of the fields corresponds to the scale factor $a(x^0)$ associated with the size of spatial slices. See~\Refe{Feinberg:1995tf} for related work.}
	
	By extending the definition~(\ref{inner}) to arbitrary complex functions $f_1(\vec \q)$, $f_2(\vec \q)$, in the sense that they are not necessarily solutions of the WDW equation, we find by integrating by parts
	\be
	\langle f_1|{\cal H}f_2\rangle= \langle {\cal H}^\dagger f_1|f_2\rangle-{\hbar^2\over 2}\int_{\partial\T}\d\Sigma \, n^i\left[{\mu f_1^*\over J}\,\nabla_i(Jf_2)-\nabla_i\Big({\mu f_1^*\over J}\Big)Jf_2\right],
	\label{hhd}
	\ee 
	where we have defined for any complex function $f$
	\be
	{\cal H}^\dagger f\equiv{1\over J}\left[-{\hbar^2\over 2}\, {J^2\over \mu}\, \nabla^2\Big({\mu\over J^2}\, Jf\Big)+vJf\right]=0\, .
	\ee
	In the right-hand side of \Eq{hhd}, the integral is a boundary term involving the normal vector $n^i$ of the boundary $\partial \T$ of the target space.  Imposing 
	\be
	{\cal H}f={\cal H}^\dagger f
	\ee
	yields a contraint for $\mu$, 
	\be
	\nabla^2\Big({\mu\over J^2}\Big)=-2 \, \nabla_i\Big({\mu\over J^2}\Big)\, {\nabla^i(Jf)\over Jf}\, .
	\ee
	Assuming that $\nabla_i({\mu/J^2})$ is not identically vanishing, this  equation leads  to a contradiction, since $Jf$ can be any function. As a result, $\nabla_i({\mu/ J^2})\equiv 0$ is satisfied, which means that
	\be
	\mu=J^2\, ,
	\label{J}
	\ee
	up to an irrelevant multiplicative positive constant we set to 1. In this case, applying \Eq{hhd} to arbitrary wavefunctions $\psi_1(\vec\q)$, $\psi_2(\vec \q)$, \ie functions annihilated by the Hamiltonian, the boundary term necessarily vanishes. As a result, we obtain
	\be
	\langle \psi_1|{\cal H}\psi_2\rangle= \langle {\cal H} \psi_1|\psi_2\rangle\, , 
	\ee 
	which is the Hermiticity constraint of the Hamiltonian.
	
	For the following reasons, it proves useful to consider rescaled wavefunctions
	\be\label{rePsi}
	\Psi\equiv J\psi \, .
	\ee
	Indeed, multiplying \Eq{wdwpsi} by $J$, they satisfy
	\be
	J{\cal H}\psi \equiv \boldsymbol{\mathcal{H}}\Psi = -{\hbar^2\over 2}\, \nabla^2\Psi+v\Psi=0\, ,
	\label{wdwPsi}
	\ee
	while the probability amplitude in \Eq{inner} reduces to
	\be
	\langle \psi_1 | \psi_2 \rangle=	( \Psi_1 | \Psi_2 ) \equiv \int_\T \prod_{k=0}^{n-1}\d \q^k \sqrt{-\gamma} \; \Psi_1(\vec \q)^*\, \Psi_2(\vec \q)\,.
	\label{innerPsi}
	\ee
	Note that in the differential equation for $\Psi$ shown in \Eq{wdwPsi}, and in the probability amplitude $(\Psi_1|\Psi_2)$, all references to the Jacobian $J$ have disappeared. {\it This shows that the quantum theory is independent of the choice of path-integral measure.} When written in terms of the dressed wavefunctions, the inner product becomes the standard quantum mechanical inner product between wavefunctions. Moreover, the boldface differential operator  $\boldsymbol{\mathcal{H}}$ defined in \Eq{wdwPsi} is the representation of the Hamiltonian acting on the Hilbert space of wavefunctions $\Psi$. Indeed,
	\be
	\begin{split}
		\langle\psi_1|{\cal H}\psi_2\rangle&=\int_\T \prod_{k=0}^{n-1}\d \q^k \sqrt{-\gamma} \; J^2\, \psi_1^*\, {\cal H}\psi_2\\
		&=\int_\T \prod_{k=0}^{n-1}\d \q^k \sqrt{-\gamma} \; \Psi_1^*\,  \boldsymbol{\mathcal{H}}\Psi_2\\
		&=(\Psi_1| \boldsymbol{\mathcal{H}}\Psi_2)\, .\esps
	\end{split}
	\ee 
	As a result, \Eq{wdwPsi} is the WDW equation for the wavefunctions $\Psi$, as it is normalized in such a way to match with the Hamiltonian.
	
	The fact that all probability amplitudes and physical observables are independent of the choice of path-integral measure, and that all quantum prescriptions related to field redefinitions are physically equivalent, closely mirrors well-known results in quantum field theory. Indeed, in quantum field theory, the classical action is invariant under field redefinitions, while there are apparent ambiguities linked with the multitude of choices for the path integral measure. However, physical observables such as S-matrix elements remain independent of these choices~\cite{Peskin:1995ev}.
	
	
	\section{Applications: Starobinsky model and JT gravity} 
	\label{appli}
	
	We have so far  considered non-linear $\sigma$-models with flat target spaces $\mathcal{T}$. While this is  restrictive, we illustrate in the present section how this framework includes relevant theories, such as inflationary models and JT gravity with a positive cosmological constant.  We provide the universal WDW equations, along with the inner products of the Hilbert spaces. 
		
	\subsection{Starobinsky model}
	
	In the gravitational action, higher derivative terms of the form $R^\xi$,  where $\xi>3/2$ is a constant and $R$ is the Ricci scalar, lead to inflationary models~\cite{Guth:1980zm, Stelle:1976gc,Stelle:1977ry}. The latter turn out to be equivalent to Einstein gravity with a minimally coupled scalar field \cite{Teyssandier:1983zz, Whitt:1984pd, Barrow:1988xh, Barrow:1988xi, Chiba:2003ir}. For simplicity, let us review  the simplest scenario, originally proposed by Starobinsky \cite{Starobinsky:1980te,Starobinsky:1983zz,Mukhanov:1981xt}, where the Einstein--Hilbert action is supplemented with an $R^2$ term.\footnote{See \Refe{Kouniatalis:2025qfz} for an earlier work on the wavefunction of the universe during inflation.}
    
    In the notations of \Eq{SE}, the action, including the generalization of the Gibbons--Hawking--York term for gravity theories with higher-derivative terms~\cite{Barth:1984jb,Casadio:2001ff,Dyer:2008hb,Balcerzak:2008bg}, reads
\begin{equation}\label{S}
		S = \int_{\mathcal{M}} \d^4x\,  \sqrt{-g}\left({R\over 2}+\frac{R^2}{16M^2}\right) +  \int_{\partial\mathcal{M}} \d^3y\, \sqrt{h} \left(1+\frac{R}{4M^2}\right)\!K  \, ,
	\end{equation}
where $M$ is a constant with the dimension of a mass. The bulk Lagrangian can be linearized in $R$ by introducing a Lagrange multiplier $\lambda$,  
	\begin{equation}
		\label{eq:Star}
		S = \int_{\mathcal{M}}\d^4x\, \sqrt{-g}\left({R\over 2}+\lambda R-4M^2 \lambda^2\right)+  \int_{\partial\mathcal{M}} \d^3y \,\sqrt{h} \,(1+2\lambda)K \, .
	\end{equation}
Indeed, using the equation of motion for $\lambda$, namely  $\lambda=R/(8M^2)$, one recovers \Eq{S} when $\lambda$ is on-shell. By redefining the Lagrange multiplier in terms of a scalar field $\mathit{\Phi}$ as
\be
\label{defphi}
\alpha \mathit{\Phi} = \ln(1+2\lambda)\, , \qquad \where\qquad \alpha=\sqrt{{2\over 3}}\, , 
\ee
and redefining the metric as 
\begin{equation}
\label{eq:Weyl}
g_{\mu\nu}= e^{-\alpha\mathit{\Phi}}\,\tilde g_{\mu\nu}\, , 
	\end{equation}
we show in \Appendix{app:Starobinsky} that the action can be cast into the Einstein-frame
	\begin{equation}
		\label{eq:star'}
		S= \int_{\mathcal{M}} \d^4x \,\sqrt{-g}\,\bigg({R\over 2}-{1\over 2}\,\partial_{\mu}\mathit{\Phi}\partial^{\mu}\mathit{\Phi} - M^2(1-e^{-\alpha\mathit{\Phi}})^2\bigg) + \int_{\partial\mathcal{M}}\d^3y \,\sqrt{h}\,K\,,
	\end{equation}
where, for notational simplicity, we have omitted tildes on all quantities.
 As announced, the model reduces to a scalar field  minimally coupled  to Einstein gravity, including the Gibbons--Hawking--York term, with a Starobinsky potential having a vanishing minimum at $\mathit{\Phi}=0$ and a positive plateau for $\mathit{\Phi}\gg 1/\alpha$.

Let us focus for now on the simpler model where spacetime is restricted to be homogeneous and isotropic. In this case, the spacetime metric and scalar field satisfy
	\begin{align}
		\label{eq:metric}
		\d s^2 &= -N(x^0)^2 \, {\d x^0}^2 +a(x^0)^2\,  \d\Omega_3^2\, ,\\
		\mathit{\Phi} &= \mathit{\Phi}(x^0)\, ,
	\end{align}
 where $\d\Omega_3^2$ is the line element on the spatial unit 3-sphere. 
 Integrating the bulk action by parts, all boundary terms cancel and the theory can be expressed in terms of the Lagrangian
\be
		L= 2\pi^2 N \Big( \!-3\, \frac{a\dot{a}^2}{N^2}+\frac{a^3}{2}{\dot{\mathit{\Phi}^2}\over N^2}+3a-M^2a^3(1-e^{-\alpha\mathit{\Phi}})^2\Big)\, ,
	\ee
	where the factor $2\pi^2$ is the unit 3-sphere volume.
By identifying $q^0=a$, $q^1=\mathit{\Phi}$, it can be brought to the form shown in \Eq{sigma}, where the target space metric and potential are 
	\begin{equation}
		\gamma_{ij}(a,\mathit{\Phi}) = 2\pi^2 \begin{pmatrix}
			-6a & 0\\
			0& a^3
		\end{pmatrix}\, , \qquad v(a,\mathit{\Phi})=2\pi^2\left( -3a+M^2a^3(1-e^{-\alpha \mathit\Phi})^2 \right).
	\end{equation}
The key point is that the minimal coupling of the scalar field to gravity implies that the target space $\mathcal{T}$ is locally flat, that is 
	\begin{equation}
		{\cal R}^i_{~jkl} = 0\, ,
	\end{equation}
where ${\cal R}^i_{\phantom{i}jkl}$ is the Riemann curvature tensor. As a result, the quantum theory can be described in terms of the universal WDW equation~\eq{wdwPsi} for the rescaled wavefunction~$\Psi({\mathrm a}, \Phi)$,
\be
\label{wdwstaro}
-{\hbar^2\over 4\pi^2}\left[-{1\over 6{\mathrm a}}{\partial^2\Psi\over \partial {\mathrm a}^2}+{1\over {\mathrm a}^3}{\partial^2\Psi\over \partial \Phi^2}-{1\over 6{\mathrm a}^2}{\partial \Psi\over \partial {\mathrm a}}\right]+v\Psi=0\, .
\ee
Moreover, the Hilbert space of solutions is equipped with the inner product
\begin{equation}
( \Psi_1 | \Psi_2 ) = 2\sqrt{6}\,\pi^2\int_0^{+\infty}\d {\mathrm a}\int_{-\infty}^{+\infty} \d \Phi\,   {\mathrm a}^2\, \Psi_1^*\, \Psi_2\, .
\end{equation}
We leave the investigation of the WDW equation~(\ref{wdwstaro}) for future work. 


	\subsection{de Sitter JT gravity}
	\label{JTex}
	
	When searching for toy models of quantum gravity, a useful strategy is to consider two-dimensional spacetimes. Since pure Einstein gravity is trivial in this case, a simple way to build an interesting model is to couple gravity to a scalar field $\mathit{\Phi}$. In particular, de Sitter JT gravity is defined by the action~\cite{Jackiw:1984je, Teitelboim:1983ux}
	\begin{equation}
		S = {1\over 2}\int \d^2x\, \sqrt{-g}\,\mathit{\Phi} (R-2)\,.
	\end{equation}
Upon variation with respect to $\mathit{\Phi}$, one obtains the de Sitter equation of motion in two dimensions, $R=2$. Even if quantization of de Sitter JT gravity  has received significant attention in recent years~\cite{Maldacena:2019cbz,Cotler:2019dcj,Cotler:2019nbi,Nanda:2023wne,Cotler:2024xzz,Buchmuller:2024ksd,Dey:2025osp,Alexandre:2025rgx}, it is not yet fully understood. Phase space aspects have also been studied in \Refs{Held:2024rmg,Alonso-Monsalve:2024oii,Buchmuller:2024ksd}. Since the operator-ordering ambiguities remain poorly understood in this context, applying our results to this model is  particularly appealing.
    
	Following \Refe{Held:2024rmg}, we parametrize the metric in ADM form
	\begin{equation}
	\label{an}
		\d s^2 = -N^2 (\d x^0)^2 + a^2(\d x^1+N_{\perp}\d x^0)^2\,,
	\end{equation}
	where $N$ is the lapse function, $N_{\perp}$ is the shift vector and $a^2$ is the spatial metric component, which are all depending on the spacetime coordinates $x^\mu$. Let us consider compact geometries, where all fields are periodic in $x^1$ with period $2\pi$. In these notations, the action reads
	\be
	\begin{split}
	S&=\int\d x^0\d x^1\, {\cal L}\, , \\
\with\qquad 	{\cal L}&= N\left(-\frac{\mathit{\Phi}''}{a}+\frac{a'\mathit{\Phi}'}{a^2}-a\mathit{\Phi} \right) - \frac{1}{N}\big(\dot{a}-(N_{\perp}a)'\big)\big(\,\,\dot{\!\!\mathit{\Phi}}-N_{\perp}\mathit{\Phi}'\big)\,,
	\end{split}
	\ee
where the dot and prime derivatives are taken with respect to $x^0$ and $x^1$, respectively. Among the fields, only $a$ and $\mathit{\Phi}$ are dynamical, since the conjugate momenta are given by
	\begin{align}
\label{pia}	\pi_a&= \frac{\partial \cal L}{\partial \dot{a}}=-\frac{1}{N}\big(\,\,\dot{\!\!\mathit{\Phi}}-N_{\perp}\mathit{\Phi}'\big)\,,\\
\label{piphi}		\pi_{\mathit{\Phi}}&= \frac{\partial \cal L}{\partial \,\,\dot{\!\!\mathit{\Phi}}}=-\frac{1}{N}\big(\dot a-(N_\perp a)'\big)\,,\\
		\pi_N&= \frac{\partial \cal L}{\partial \dot N}=0\,,\\
		\pi_{N_\perp}&= \frac{\partial \cal L}{\partial \dot N_\perp}=0\, .
	\end{align}
		Applying a Legendre transformation, the Hamiltonian reads
	\begin{equation}
\int \d x^1\big(	\pi_a\dot{a} + \pi_{\mathit{\Phi}}\,\,\dot{\!\!\mathit{\Phi}} - {\cal L}\big)\equiv \int \d x^1 (H_0+P) \,,
	\end{equation}
	where we have defined
	\begin{align}
    \label{eq:Hconstraint}
		{H_0\over N}&= -\pi_a\pi_\mathit{\Phi} + a\mathit{\Phi} + \frac{\mathit{\Phi}''}{a}- \frac{a'\mathit{\Phi}'}{a^2} \equiv-{\delta S\over \delta N} \,,\\
        \label{eq:Pconstraint}
		{P\over N_\perp}&= \pi_{\mathit{\Phi}}\mathit{\Phi}' - \pi_a'a \equiv-{\delta S\over \delta N_\perp}\,.
	\end{align}
Thanks to the last identities, the classical system obeys the Hamiltonian and momentum constraints on shell,
	\be
	\label{Hmom}
		{H_0\over N}=0\,, \qquad {P\over N_\perp}=0\,.
	\ee
	
	In \Appendix{app:minisuperspace}, we review the fact that invariance of the theory under diffeomorphisms in two dimensions allows to choose a coordinate system in \Eq{an} for which all fields and their conjugate momenta are independent of $x^1$, except in some subspaces of the phase space~\cite{Louis-Martinez:1993bge,Buchmuller:2024ksd}. In our work, ignoring these particular subspaces, the minisuperspace approach appears to be exact rather than approximate.  When all fields are homogeneous, the momentum constraint is trivial and the Hamiltonian constraint can be recast in terms of $H=\int \d x^1 H_0=2\pi H_0$, 

	\begin{equation}
		{H\over N} = 2\pi\!\left(-{\dot a\,\,\dot{\!\!\mathit{\Phi}} \over N^2} + a\mathit{\Phi}\right) \!=0\,,
	\end{equation} 
where all fields are independent of $x^1$. Denoting $q^0=a$, $q^1=\mathit{\Phi}$, we can write the Lagrangian $L=\int \d x^1 \,{\cal L}=2\pi{\cal L}$ as in \Eq{sigma}, with the target-space metric and potential given by  
	\be
		\gamma_{ij}(a,\mathit{\Phi}) = - 2\pi\begin{pmatrix}
			0& 1\\
			1& 0
		\end{pmatrix} ,\qquad v(a,\mathit{\Phi}) = 2\pi\, a\mathit{\Phi}\,.
	\ee
Since $\gamma_{ij}$ is constant, with determinant  $\gamma<0$, the target space $\T$ is flat and Lorentzian. Our treatment of the ordering ambiguity can therefore be applied, and the universal WDW equation~\eqref{wdwPsi} reads
\begin{equation}
\label{eq:WDWJT}
{\hbar^2\over 2\pi}\,\frac{\partial^2\Psi}{\partial {\rm a} \partial\Phi}+2\pi\, {\rm a\Phi}\, \Psi=0\, ,
\end{equation}
with inner product~(\ref{innerPsi})	
\begin{equation}
	(\Psi_1|\Psi_2) = 2\pi \int_0^{+\infty}\d {\rm a}  \int_{-\infty}^{+\infty}\d\Phi \;\Psi_1^*\Psi_2\, . 
\end{equation}

This quantum theory differs from those commonly found in the literature. In particular, it contrasts with the approach based on Henneaux's factor ordering~\cite{Henneaux:1992ig,Nanda:2023wne,Buchmuller:2024ksd}, where one first solves both constraints classically and then imposes them as first-order operator conditions on the wavefunction~\cite{Henneaux:1992ig}. More recently, the authors of \Refe{Held:2024rmg} have considered the WDW equation \eqref{eq:WDWJT}.\footnote{In the notations of~\Refe{Held:2024rmg}, our fields $a$, $\mathit{\Phi}$ are denoted $a/\sqrt{2\pi}$, $\mathit{\Phi}/\sqrt{2\pi}$. As a result, no factor $2\pi$ appears in their WDW equation.} However, they adopt a definition of the inner product based on group averaging, which differs from ours.
	
	
\section{Conclusion}
	\label{compa}

In this work, we have resolved the ordering ambiguity that arises in the canonical quantization of the Hamiltonian for any minisuperspace model with an arbitrary number of degrees of freedom and quadratic kinetic term, provided the target space is flat and the Universe is closed. To each admissible operator ordering of the Hamiltonian corresponds a specific WDW equation for the wavefunction of the Universe. The criterion for an ordering to be admissible is the existence of a choice of path-integral measure such that the corresponding wavefunction path integral solves the associated WDW equation.

The regularization scheme used to define the path integral is skeletonization, which amounts to integrating over field values at discrete cosmological times. However, the fields appearing in the action and those obtained after field redefinitions generally lead to distinct path-integral measures, since integrating over the values of the former is not equivalent to integrating over the values of the latter. To systematically analyze all possible measures, we first consider the skeletonization of canonical fields, which is always possible when the target space is flat. 
At each discrete time step, we then dress the integration over the values of the canonical fields by the Jacobian relating them to those of the redefined fields. 
As a result, any path-integral measure is fully characterized by a Jacobian, which is a real, positive function. The set of consistent operator orderings in the quantum Hamiltonian therefore forms a tiny, though infinite, subset of all possible orderings, in one-to-one correspondence with the Jacobians associated with field redefinitions.

The resulting WDW equations, namely those compatible with the Feynman path-integral viewpoint, lead to equivalent quantum theories. This equivalence is established by deriving the appropriate Hilbert-space inner product for each operator ordering through the requirement of Hermiticity of the quantum Hamiltonian. Using the correct  inner products, the observables turn out to be independent of the operator ordering, or equivalently of the path-integral measure. These inner products also possess several important properties. First, they are genuinely positive definite, unlike the Klein--Gordon ``norm'' that has often been employed in the literature since the original work on the WDW equation~\cite{DeWitt:1967yk}. Second, at least in some simple minisuperspace models~\cite{Kehagias:2021wwr}, our norm $|\Psi|^2$ allows one to recover the classical cosmological evolution in the $\hbar\to0$ limit. Since $|\Psi|^2$ does not involve an explicit notion of time, our formalism provides a natural resolution of the problem of time in quantum gravity.

The restriction of our analysis to the case where the minisuperspace (the target space $\T$ of the $\sigma$-model Lagrangian) is flat can be traced to the following reason. When $\T$ is curved, one can no longer work in Minkowski coordinates, and the metric $\gamma_{ij}$ necessarily depends on the coordinates $\vec \q$. Proceeding as in \Eqs{eq}--(\ref{eb}), the expansion of the integrand in the analogue of \Eq{eq} then involves inverse powers of $\varepsilon$, which makes term-by-term integration illegitimate. As a consequence, an analogue of \Eq{eq1} cannot be derived within this framework. It is therefore of primary importance to develop an alternative strategy to generalize our results to curved target spaces $\T$. In such cases, as often advocated, the WDW equation is expected to involve a curvature-dependent correction to the potential term. Needless to say that the class of models with curved target spaces is much bigger. It includes supergravity models of chaotic and Starobinsky-type inflation where the scalar-field manifold is curved, no-scale supergravity models, multi-field inflationary models, string modular cosmology where the moduli space is curved, anisotropic models such as the interior of a black hole, and many others. 

We have applied our results to several interesting models, including JT gravity and the Starobinsky model. In the latter case, however, we have quantized the theory in its formulation where a scalar field is coupled to Einstein gravity. This form is not obtained directly by a field redefinition of the original Lagrangian containing an $R^2$ term. As a result, it is not \apriori clear whether a direct quantization of the $R^2$ theory would lead to the same quantum theory. Clarifying this point would certainly be worthwhile.

Finally, let us place our results in the context of the existing literature. The WDW equation~(\ref{wdwPsi}) and the inner product~(\ref{innerPsi}), valid for flat target spaces, already appear in \Refe{Christodoulakis:1984gp}. In that work, the authors perform a field redefinition to express the Lagrangian in terms of canonical fields, for which no ordering ambiguity arises in canonical quantization, and subsequently revert to the original variables. However, the ambiguities that emerge from the path-integral perspective are not addressed. In particular, the alternative forms of the WDW equation~(\ref{wdwpsi2}) and of the inner products~(\ref{inner}),~(\ref{J}), as well as their equivalence, are not discussed. The derivation of the WDW equation from the path-integral viewpoint has also been considered previously. In \Refe{Halliwell}, however, several issues related to the gauge fixing of time reparametrization in the path integral arise, and the determination of the Hilbert-space inner product is not addressed. Further discussion of these issues can be found in \Refe{WDW1D}.


\begin{appendices}
\makeatletter
\DeclareRobustCommand{\@seccntformat}[1]{%
  \def\temp@@a{#1}%
  \def\temp@@b{section}%
  \ifx\temp@@a\temp@@b
  \appendixname\ \thesection:~~%
  \else
  \csname the#1\endcsname~~%
  \fi
} 
\makeatother


\section{Starobinsky action with boundary term}
\label{app:Starobinsky}
\renewcommand{\theequation}{A.\arabic{equation}}

In this appendix, we review the main steps showing that the classical Starobinsky model can be recast as Einstein gravity minimally coupled to a scalar field, explicitly keeping the boundary term appearing in \Eq{S}.

Our starting point is \Eq{eq:Star}, which involves the Lagrangian multiplier $\lambda$. Using the definition in \Eq{defphi}, we split the action into two parts as
\begin{align}
S &= S_{\mathcal{M}} + S_{\partial\mathcal{M}}\, ,\\
	\where\qquad	S_{\mathcal{M}} &= \int_{\mathcal{M}}\d^4x\, \sqrt{-g}\left(e^{\alpha\mathit{\Phi}}\frac{R}{2}-M^2 \left(e^{\alpha\mathit{\Phi}}-1\right)^2\right),\\
        \label{eq:Sbdy}
        S_{\partial\mathcal{M}}&=\int_{\partial\mathcal{M}} \d^3y \,\sqrt{h} \,e^{\alpha\mathit{\Phi}}K \, .
\end{align}
In what follows, we rewrite $S$ in the Einstein frame by defining
\begin{equation}
		g_{\mu\nu}= e^{-\alpha\mathit{\Phi}}\,\tilde g_{\mu\nu}\, , 
\end{equation}
and we denote with tildes all quantities computed with respect to the metric $\tilde g_{\mu\nu}$. 
Indeed, the Ricci scalar satisfies
\begin{equation}
    R =e^{\alpha\mathit{\Phi}}\big(\tilde R + 3\alpha \tilde\nabla_{\mu}\tilde\nabla^{\mu}\mathit{\Phi} - \tilde\nabla_{\mu}\mathit{\Phi}\tilde\nabla^{\mu}\mathit{\Phi}\big)\, ,
\end{equation}
and the bulk action reads
\begin{equation}
\label{eq:StildeM}
    S_{\mathcal{M}} = \int_{\mathcal{M}}d^4x \,\sqrt{-\tilde g}\left(\frac{\tilde R}{2}+\frac{1}{\alpha}\,\tilde\nabla_{\mu}\tilde\nabla^{\mu}\mathit{\Phi}-\frac{1}{2}\,\tilde\nabla_{\mu}\mathit{\Phi}\tilde\nabla^{\mu}\mathit{\Phi}-M^2\left(1-e^{-\alpha\mathit{\Phi}}\right)^2\right).
\end{equation}
In \Eq{eq:Sbdy}, the trace of the extrinsic curvature is defined as 
\be
K=h^{\mu\nu}\nabla_{\mu}n_{\nu}\,,
\ee
with $n^{\mu}$ the unit vector normal to the spacelike boundary $\partial\mathcal{M}$, and $h_{\mu\nu}$ the induced metric on $\partial\mathcal{M}$:
\be
h_{\mu\nu}=g_{\mu\nu}-n^{\sigma}n_{\sigma}n_{\mu}n_{\nu}\, , \qquad \where\qquad n^{\mu}n_{\mu}=-1\, .
\ee
Normalization of the unit vector in the Einstein frame requires 
\begin{equation}
    n_{\mu} = e^{-\frac{\alpha\mathit{\Phi}}{2}}\,\tilde n_{\mu}\, ,
\end{equation}
which leads to 
\be
h_{\mu\nu}=e^{-\alpha\mathit{\Phi}}\,\tilde h_{\mu\nu}\, ,
\ee
while the Christoffel symbols satisfy
\begin{equation}
    \Gamma^{\lambda}_{\mu\nu} = \tilde\Gamma^{\lambda}_{\mu\nu}+\partial_{\{\mu}\mathit{\Phi}~\delta^{\lambda}_{\nu\}}-\partial_{\sigma}\mathit{\Phi}\,\tilde g^{\sigma\lambda}\tilde g_{\mu\nu}\, .
\end{equation}
A straightforward computation then yields
\begin{equation}
\label{eq:StildedM}
    S_{\partial\mathcal{M}}= \int_{\partial\mathcal{M}}d^3y \,\sqrt{\tilde h}\left(\tilde K-\frac{1}{\alpha}\,\tilde n^{\mu}\tilde\nabla_{\mu}\mathit{\Phi}\right).
\end{equation}
By Stokes' theorem, we have 
\begin{equation}
     \int_{\mathcal{M}}d^4x \,\sqrt{-\tilde g}\,\tilde\nabla_{\mu}\tilde\nabla^{\mu}\mathit{\Phi} = \int_{\partial\mathcal{M}}d^3y \,\sqrt{\tilde h}\,\tilde n^{\mu}\tilde\nabla_{\mu}\mathit{\Phi}\, ,
\end{equation}
so that the second terms in \Eq{eq:StildeM} and \Eq{eq:StildedM} cancel. As a result, the full action reduces to 
	\begin{equation}
		S= \int_{\mathcal{M}} \d^4x \,\sqrt{-\tilde g}\left({\tilde R\over 2}-{1\over 2}\,\tilde\nabla_{\mu}\mathit{\Phi}\tilde\nabla^{\mu}\mathit{\Phi} - M^2(1-e^{-\alpha\mathit{\Phi}})^2\right) + \int_{\partial\mathcal{M}}\d^3y \,\sqrt{\tilde h}\,\tilde K\,,
\end{equation}
which is \Eq{eq:star'}, where the tildes have been dropped for notational simplicity.


\section{de Sitter JT gravity in minisuperspace}
\label{app:minisuperspace}
\renewcommand{\theequation}{B.\arabic{equation}}

In de Sitter JT gravity, the minisuperspace approach is not an approximation but rather a gauge choice~\cite{Louis-Martinez:1993bge,Buchmuller:2024ksd}, except in some subspaces of the phase space~\cite{Held:2024rmg}. We clarify this statement in the following appendix.

Thanks to diffeomorphism invariance in two dimensions, let us choose 
a coordinate system in~\Eq{an} such that 
\begin{equation}
\label{hypo}
a'=0\, , \qquad \pi_\mathit{\Phi}'=0\, .
\end{equation}
In this case, the quantity 
\begin{equation}
\label{Ceq}
C = \pi_{\mathit{\Phi}}\left(a\pi_a-\mathit{\Phi}\,\pi_{\mathit{\Phi}}\right)
\end{equation}
satisfies
\begin{equation}
\label{c'}
C' = -\frac{\pi_{\mathit{\Phi}}}{N_{\perp}}P =0\, ,
\end{equation}
where the last identity follows from the momentum constraint in~\Eq{Hmom}. Moreover, the Hamiltonian constraint in~\Eq{Hmom} becomes
\begin{equation}
\label{HH}
\frac{H_0}{N}= -\frac{C}{a} -\frac{\pi_{\mathit{\Phi}}^2\mathit{\Phi}}{a}+a\mathit{\Phi}+\frac{\mathit{\Phi}''}{a}=0\, .
\end{equation}
Defining 
\begin{equation}
\mathcal{K}^2 = a^2-\pi_{\mathit{\Phi}}^2\, , 
\end{equation}
\Eq{HH} can be recognized as a  second-order linear differential equation,
\begin{equation}
\label{eqq}
\mathit{\Phi}'' +\mathcal{K}^2\mathit{\Phi} = C\, ,
\end{equation}
since $\mathcal{K}^2$ and $C$ depend only on $x^0$. The generic solution is 
\be
\label{solu}
\mathit{\Phi} =\left\{\begin{array}{ll}
\dis  \kappa_1\cos\!\big(\mathcal{K}x^1\big) + \kappa_2 \sin\!\big(\mathcal{K}x^1\big)+\frac{C}{\mathcal{K}^2}\, ,  & \quad \when\qquad \espD \mathcal{K}(x^0)\neq 0\, , \\
\dis  \kappa_3 +\kappa_4\, x^1+ C\, {{x^1}^2\over 2}\, , &\quad  \when\qquad \mathcal{K}(x^0)=0\, , \end{array}\right. 
\ee
where $\kappa_1,\dots, \kappa_4$ are  fonctions of $x^0$. Owing to the $2\pi$-periodicity of $\mathit{\Phi}$ with respect to $x^1$ (see below~\Eq{an}), we have $\kappa_4(x^0)=C(x^0)=0$ in the second line of \Eq{solu}. Moreover, two cases may arise:
\begin{itemize}
\item When $\mathcal{K}(x^0)$ is a constant, equal  to a non-vanishing integer, $\mathcal{K}(x^0)\equiv k\in\Z\backslash\{0\}$, the scalar $\mathit{\Phi}$ can be non-homogeneous.

\item In all other cases, one must have $\kappa_1(x^0)=\kappa_2(x^0)\equiv 0$ in the first line of \Eq{solu}, implying that $\mathit{\Phi}$ is homogeneous.
\end{itemize}
When $\mathit{\Phi}$ is homogeneous, since $a$ and $\pi_\mathit{\Phi}$ are homogeneous from the outset (see \Eq{hypo}), it follows that $\pi_a$ (see~\Eq{Ceq}), $N$ (see \Eq{pia}), and $N_\perp'$ (see~\Eq{piphi}) are also homogeneous. Consequently, $N_\perp$ must take the form $N_\perp = \kappa_5 x^1 + \kappa_6$, where $\kappa_5,\kappa_6$ are functions of $x^0$. Periodicity in $x^1$ then imposes $\kappa_5 = 0$, so that $N_\perp$ is homogeneous.

To conclude, in a coordinate system where $a'=\pi_\mathit{\Phi}'=0$, 
the minisuperspace approach is exact, except in the subspaces of the phase space where $\sqrt{a^2-\pi_{\mathit{\Phi}}^2}$ is a non-zero integer.\footnote{For completeness, let us mention that extra non-homogeneous solutions to the local classical equations of motions exist. Starting from a configuration where $\sqrt{a^2-\pi_{\mathit{\Phi}}^2}\equiv k\in\N\backslash\{0\}$, one can switch on a parameter $\beta$ that deforms the boundary conditions to $(x^0,x^1)\sim (x^0+2\pi\beta,x^1+2\pi)$, thereby yielding geometries interpreted as multicovers of de Sitter spacetime~\cite{Held:2024rmg}.}

\end{appendices}




	
\section*{Acknowledgements}
	We thank Thomas Mertens, Bruno de S. L. Torres and Vincent Vennin for useful discussions. N.T. thanks the Ecole Polytechnique for its hospitality. V.F. acknowledges financial support from the European Research Council (grant BHHQG-101040024). Views and opinions expressed are however those of the authors only and do not necessarily reflect those of the European Union or the European Research Council. Neither the European Union nor the granting authority can be held responsible for them.
	\vspace{.5cm}

	
	

\end{document}